\documentclass{jpp}

\usepackage{subeqnarray}
\usepackage{graphicx}
\usepackage{subfig}
\usepackage{color}
\usepackage{amsbsy, amsgen,upmath}
\usepackage{amsmath,amssymb,verbatim}
\usepackage[round]{natbib}

\newcommand{\llangle}{\ensuremath{\langle \! \langle }}
\newcommand{\rrangle}{\ensuremath{\rangle \! \rangle }}

\newcommand{\h}[1]{\ensuremath{\widehat{#1}}}
\newcommand{\bb}[1]{\ensuremath{{\boldsymbol{#1}}}}

\newcommand{\za}[1]{\ensuremath{\langle{#1}\rangle}}
\newcommand{\ba}[1]{\ensuremath{\llangle{#1}\rrangle}}

\newcommand{\pb}[2]{\ensuremath{\left\{#1,#2\right\}}}
\newcommand{\bdot}{\ensuremath{\, \boldsymbol{\cdot}\,}}
\newcommand{\btimes}{\ensuremath{\, \boldsymbol{\times}\,}}

\newcommand{\del}{\ensuremath{\boldsymbol{\nabla}}}

\newcommand{\eq}[1]{(\ref{#1})}

\newcommand{\rr}[1]{\ensuremath{{\mathrm{#1}}}}

\newcommand{\dd}{\ensuremath{\operatorname{d}\!}}

\newcommand{\THd}{\ensuremath{\h{\delta}}}
\newcommand{\jpi}{\ensuremath{\upi }}
\newcommand{\ji}{\ensuremath{\rr{i} }}
\newcommand{\je}{\ensuremath{\rr{e} }}

\begin{document}

\title{On Nonlocal Energy Transfer via Zonal Flow in the Dimits Shift}

\author[D.\:A.\:St-Onge]{Denis A.\:St-Onge$^{1,2}$ \thanks{E-mail address for correspondence: dstonge@princeton.edu}}

\date{\today}
\affiliation{$^1$Department of Astrophysical Sciences, Princeton University, Princeton, New Jersey 08544, USA\\[\affilskip]
$^2$Princeton Plasma Physics Laboratory, P.O. Box 451, Princeton, NJ 08543, USA}

\maketitle

\begin{abstract}
The two-dimensional Terry-Horton equation is shown to exhibit the Dimits shift when suitably modified to capture both the nonlinear enhancement of zonal/drift-wave interactions and the existence of residual Rosenbluth-Hinton states. This phenomenon persists through numerous simplifications of the equation, including a quasilinear approximation as well as a four-mode truncation. 
It is shown that the use of an appropriate adiabatic electron response, for which the electrons are not affected by the flux-averaged potential, results in an $\bb{E}\btimes\bb{B}$ nonlinearity that can efficiently transfer energy nonlocally to length scales on the order of the sound radius.  The size of the shift for the nonlinear system is heuristically calculated and found to be in excellent agreement with numerical solutions.  The existence of the Dimits shift for this system is then understood as an ability of the unstable primary modes to efficiently couple to stable modes at smaller scales, and the shift ends when these stable modes eventually destabilize as the density gradient is increased. This nonlocal mechanism of energy transfer is argued to be generically  important even for more physically complete systems.
\end{abstract}

\section{Introduction}

The Dimits shift is the nonlinear upshift of the critical temperature gradient for the onset of turbulent transport witnessed in simulations of collisionless tokamak plasmas~\citep{Dimits}. This shift results from the shearing away of turbulent radial streamers by poloidal zonal flows that are generated from the so-called secondary instability~\citep{rogers}. The shearing of radial streamers leads to fine-scale structure, which is subsequently damped. The zonal flows are then able to persist on a longer time scale than are the turbulent eddies, since they are not Landau-damped. 
These residual zonal flows are called Rosenbluth-Hinton states~\citep{rosenbluth}, named for those who were the first to show this property.

However, as the temperature gradient is further increased, the system eventually experiences a tertiary instability~\citep{rogers}, giving way to turbulent transport. 
While the qualitative aspects of the Dimits shift are understood, there is yet no complete theory that can simultaneously predict the basic features of the shift, such as its size and dependence on various physical parameters.
  Understanding the essential aspects of this phenomenon is critical, as it is known that zonal flows have the ability to suppress turbulence in both physical systems~\citep{biglari1990,carter,schaffner} and simplified ones~\citep{Ricci_PRL}. Such a mechanism is one candidate for explaining the L-H transition seen in tokamaks~\citep{Burrell}. Being able to predict the saturated level of zonal flows and their effects on turbulence is crucial for the enhancement of plasma confinement~\citep{terry2000}.

The Dimits shift was first witnessed in gyrokinetic simulations of electrostatic toroidal plasmas~\citep{dimits1998b}, and was eventually demonstrated in nonlinear gyrofluid models with Landau-fluid closures~\citep{beer98}. This culminated in a landmark comparative study of various gyrokinetic and gyrofluid codes by~\citet{Dimits}.
The shift has also been captured in some fluid models using a variety of simplifications.  One such model is the minimal two-field ion-temperature-gradient (ITG) system, which retains both the ion continuity equation and an equation for the perpendicular ion temperature~\citep{ottaviani1, kolesnikov1, kolesnikov2}. Another important model in the study of the zonal-flow/drift-wave interaction has been the two-field Hasegawa-Wakatani model (studied separately by~\citealt{numata} and~\citealt{Farrell2}), which is a system that includes both ion-density-gradient drift waves and non-adiabatic electron effects.

The first quantitative study on the secondary instability was by performed by~\citet{Cowley_ITG}, who considered a simplified three-field model of the slab ITG mode. It was found that the primary slab instability resulted in elongated eddies (`streamers') that were not well accounted for by mixing length arguments, and that these eddies were themselves subject to secondary instabilities that lead to the fragmentation of the primary streamers. Both the secondary and tertiary instabilities of the toroidal ITG mode were later studied by~\citet{rogers} using both numerical gyrokinetic simulations and analytical study of a simplified two-field ITG model with finite-Larmor-radius effects. It was found that the nature of the adiabatic electron response resulted in an asymmetry in the secondary and tertiary instabilities, the former being of Kelvin-Helmholtz type, whereas the latter was generally much weaker, requiring a zonal component of perpendicular temperature for instability. This was further elaborated in~\citet{jenko_2000}, where it was shown that the differences between the electron-temperature-gradient (ETG) and ITG modes were due to the differences between the response of the adiabatic species; the adiabatic electron response for ITG modes led to an additional $\bb{E}\btimes\bb{B}$ nonlinearity which suppressed the Kelvin-Helmholtz mode in the tertiary instability, a feature that is absent in the ETG mode.

Since these pioneering works, further progress has been made into understanding the behaviour 
of the shift using both numerical simulation and analytical techniques, with much work focusing on the two-field  Hasegawa-Wakatani and ITG systems.  \citet{numata} were able to show a sharp transition between steady turbulent states and relatively quiescent states dominated by zonal flows.
 However, their simulations had similar levels of viscous damping on both the drift-wave and zonal modes, which prevented them from making a connection to the steady Rosenbluth-Hinton states. They also made a prediction on the boundary of the shift using a stability analysis based on the Kelvin-Helmholtz instability using a simplified flow profile, though the result over-predicts the shift's size.
\citet{Farrell2} used statistical closure techniques to clearly separate the dynamics of the zonal flows and of the drift-wave modes using a closure called the second-order cumulant expansion, or CE2. Here, the equations of motion are separated into zonally averaged and fluctuating parts, with eddy-eddy self-interaction terms being dropped in the fluctuation evolution equation. (This closure is equivalent to a quasilinear system when ergodicity is assumed.) They were able to demonstrate that the essential physical effects are captured even without the eddy-eddy self-interactions. 
As the drift-wave and zonal equations are also separated, they were able to use a physically relevant frictional zonal damping (as would result from ion-ion collisions, see~\citealt{lin}) that was distinct from the drift-wave damping.
However, their simulations were also stochastically forced to imitate the homogeneous turbulence lost upon neglecting the eddy-eddy self-interactions. This approach is invalid when studying the Dimits shift, however, as there is no turbulence in this regime, thus they were not able to observe a Dimits shift. It remains unclear as to whether the zonal flows in the Hasegawa-Wakatani model can entirely quench drift-wave turbulence as seen in the collisionless simulations of~\citet{Dimits}. 

Analytical work has also been done using the minimal two-field toroidal ITG model. ~\citet{kolesnikov1,kolesnikov2} were able to calculate the Dimits shift of a four-mode truncated system (4MT) using the tools of dynamical systems. However, the size of the shift was found to be strongly dependent on the truncation number of the system. Additionally, the system failed to saturate beyond the Dimits shift. This model has also been studied under the CE2 framework by~\citet{StOnge1}, in which the effect of discrete-particle noise on the onset of zonostrophic instability was studied.  Even for such a simple two-field system, the CE2 results in a system of five quantities (two zonal-averaged fields and three independent components of a covariance tensor with the off-diagonal component being complex) and even the simplest of linear calculations can quickly become tedious.  
 In addition, both the Hasegawa-Wakatani and minimal two-field ITG systems are non-normal, and as a result can be made to exhibit subcritical turbulence, a phenomenon that can obfuscate the essential physics of drift waves and zonal flows.

What, then, is the simplest model that exhibits the Dimits shift with a minimal amount of physics? To be successful, the model must
\begin{enumerate}
\item contain a nonlinear boost in the interaction between zonal flows and drift waves, leading to an increase in energy transfer between these modes and a quenching of turbulence through the shearing of eddies (this `boost' is made specific in section~\ref{Review});
\item allow for the existence of steady states consisting purely of zonal modes. For these solutions to be proper steady states, zonal modes would necessarily be linearly undamped;
\item exhibit a finite Dimits shift. In other words, given a sufficient amount of linear drive, steady zonal states must give way to turbulence leading to a finite amount of turbulent transport;
\item have the ability for saturation beyond the end of the Dimits shift without the need for zonal damping. Otherwise, saturated states may depend on the amount of zonal damping, in contrast with the original numerical studies of~\citet{Dimits}.
\end{enumerate}
The goal of this article is to show that the Terry-Horton equation can be suitably modified to exhibit all four traits. The resulting equation, henceforth referred to as the modified Terry-Horton equation (mTHE), can be used to gain insight into the fundamental aspects of the Dimits shift. In particular, it is shown that the use of an appropriate adiabatic electron response, for which the electrons are not affected by the flux-averaged potential, results in an $\bb{E}\btimes\bb{B}$ nonlinearity that can transfer energy nonlocally in $k$-space via zonal shearing, depositing energy from scales at $L \gg \rho_\rr{s} $ to scales at $L \sim \rho_\rr{s}$ (here, $\rho_\rr{s}$ is the sound radius, defined in section~\ref{Review}.) This energy transfer is in stark contrast to the usual Kolmogorov-type cascade which is scale-by-scale. By using some simple considerations of this nonlocal transfer, the existence of the Dimits shift can be understood in terms of efficient coupling to small-scale stable modes, and the size of the shift depends on when these small-scale modes destabilize. This effect is a generic property for systems with this type of adiabatic  electron response, and should be relevant even when adopting a more physically complete framework (e.g. gyrokinetics). The concepts presented here should then hopefully serve to connect the previous work of~\citet{rogers} with the more recent ideas of mode coupling (see, for example,~\citealt{Hatch_PRL,Makwana2}).

Before proceeding, this article focuses solely on the \emph{collisionless} Dimits shift, where the zonal flows are linearly undamped. 
In physical systems, some amount of zonal damping must of course be present. Even so, the underlying mechanisms present in idealized models, such as the one presented herein, which capture the fundamental aspects of the Dimits shift, should remain important when additional physical effects, such as zonal damping, are included. Remarkably, it has been noted recently by~\citet{Mikkelsen} that the Dimits shift can persist even in more physically complete systems. Thus, as the mechanisms that underlie the Dimits shift appear to be robust, studying them in isolation by using a minimal model can pinpoint which effects are truly necessary for the shift to exist. This is precisely the point of view adopted in this paper.

The remainder of this article is organized as follows. In Sec.~\ref{Review} I describe the historically important models that lead up to the mTHE, which is described in Sec.~\ref{sMTHE}. In Sec.~\ref{numerics} I demonstrate by direct numerical simulation that the mTHE exhibits the Dimits shift.  Analytical results are presented in Sec.~\ref{analytics}, starting with analysis of the 4MT, followed by a heuristic calculation of the size of the Dimits shift for the nonlinear system. This is done in Sec.~\ref{DScalc}.
 Finally, the work is summarized in Sec.~\ref{discussion}.

\section{Review} \label{Review}

\subsection{The (Modified) Hasegawa-Mima Equation}

The paradigmatic model for density-gradient-driven drift waves is the Hasegawa-Mima equation (HME), which captures advection of both the background and perturbed ion gyrocenter densities caused by fluctuations of the electrostatic potential on a local segment of the outboard midplane. The system is two-dimensional in that it neglects any toroidal variation in the electrostatic potential (i.e. $k_\parallel \ll k_\perp$); thus the fields can be described with just radial and poloidal coordinates.

 The HME is given by
\begin{equation} \label{HME}
\partial_t \zeta + \bb{v} \bdot \del\zeta - \beta \partial_y \varphi  = 0,
\end{equation}
where $\bb{v} = (-\partial_y \varphi, \partial_x \varphi)$ is the $\bb{E}\btimes \bb{B}$ drift velocity, $\beta \doteq a/ L_n$ parameterizes the density gradient, $L_n^{-1} \doteq -\dd\, \ln n_0 / \dd r$ is the background density gradient scale length, 
\begin{equation} \label{poisson}
\zeta \doteq - \delta n_\rr{i} = (\nabla_\perp^2-1) \varphi 
\end{equation}
 is the generalized vorticity derived from the gyrokinetic Poisson equation assuming an adiabatic electron response $\delta n_\rr{e} = \varphi$, $\delta n_\rr{e}$ is the perturbed electron density, and $\varphi \doteq (e \phi / T_\rr{e})(a / \rho_\rr{s})$ is the dimensionless perturbation to the  electrostatic potential. 
 Here, $\delta n_\rr{i}$ is the perturbed ion gyrocenter density normalized to the background density, $e$ is the unit charge, $\phi$ is the perturbed electrostatic potential, and $T_\rr{e}$ is the background electron temperature.
   Throughout this paper, time and space are normalized by $a/ c_\rr{s}$ and $\rho_\rr{s}$, respectively, where $a$ is the minor radius, $\rho_\rr{s} \doteq c_\rr{s} / \omega_\rr{ci}$ is the sound radius, and $c_\rr{s} \doteq (ZT_\rr{e}/m_\rr{i})^{1/2}$ is the sound speed. Coordinates are in a local Cartesian grid where $x$ represents the radial coordinate and $y$ represents the poloidal coordinate.  The operator $\nabla_\perp$ denotes the gradient perpendicular to the magnetic field and acts in the $x$ and $y$ direction for this particular setup. For any function $f$,
\begin{align}
\bb{v}\bdot \del f &= \frac{\partial \varphi}{\partial x}\frac{\partial f}{\partial y} - \frac{\partial \varphi}{\partial y}\frac{\partial f}{\partial x}
\doteq \pb{\varphi}{f}
\end{align}
which defines the Poisson bracket  $\{ \ldotp, \ldotp\}$. This nonlinearity, which arises from the $\bb{E}\btimes\bb{B}$ drift, consist of two components: the ion-polarization or vorticity nonlinearity $\pb{\varphi}{\nabla_\perp^2 \varphi}$ and the $\bb{E}\btimes\bb{B}$ nonlinearity $\pb{\varphi}{\delta n_\rr{e}}$.
While the ion-polarization nonlinearity has a specific form, the $\bb{E}\btimes\bb{B}$ nonlinearity can vary depending on which physical effects are included in the electron response.
 As the Poisson bracket is anticommutative and bilinear, it follows that for purely adiabatic electron response,
\begin{align} \label{HMpol}
 \pb{\varphi}{ \zeta}=  \pb{\varphi}{ \nabla_\perp^2 \varphi} - \pb{\varphi}{ \delta n_\rr{e}}  =  \pb{\varphi}{ \nabla_\perp^2 \varphi} - \pb{\varphi}{ \varphi}  = \pb{\varphi}{ \nabla_\perp^2 \varphi}.
\end{align}
 Thus, the $\bb{E}\btimes\bb{B}$ nonlinearity does not appear in the HME.

The nonlinear interaction in the HME conserves two quadratic invariants. These are the energy density
\begin{equation}
\mathcal{E} \doteq \frac{1}{L_xL_y} \int_0^{L_x}\dd x \int_{0}^{L_y}\dd y \, \frac{1}{2}\left[(\nabla_\perp \varphi)^2 + \varphi^2\right],
\end{equation}
and the generalised enstrophy density
\begin{equation}
\mathcal{Z} \doteq \frac{1}{L_xL_y} \int_0^{L_x}\dd x \int_{0}^{L_y}\dd y \, \frac{1}{2} \zeta^2.
\end{equation}
Because the HME is a two-dimensional equation with two nonlinearly conserved quantities, the system experiences a dual cascade where injected energy at some intermediate scale simultaneously causes a flow of energy to larger scales and a flow of enstrophy to smaller scales. Note that conservation of the generalized enstrophy $\mathcal{Z}$ is a direct result of the nonlinear interaction being in the form of a Poisson bracket and does not depend on the specific form of $\zeta$ given by \eq{poisson}. This fact is important for the Terry-Horton equation, which will be discussed in Sec.~\ref{sTHE}.

  One shortcoming of the HME is that it does not capture the correct zonal flow physics as seen in more complete gyrokinetic simulations~\citep{Dorland_1993,Hammett_developments}. This shortcoming stems from how one derives the adiabatic electron response. For the HME, this is done via parallel force balance in the electron momentum equation:
 \begin{equation}\label{elemom}
  -e n_\rr{e} E_\parallel - \nabla_\parallel P_\rr{e} =0.
 \end{equation}
 where $E_\parallel$ is the electric field parallel to the magnetic field, $n_\rr{e}$ and $P_\rr{e}$ are respectively the electron density and pressure, and $\nabla_\parallel$ is the gradient along the magnetic field ($\nabla_\parallel \rightarrow \partial_z$ for HME).
Assuming isothermal electrons, \eq{elemom} leads to
\begin{equation}
\nabla_\parallel \left(\varphi - \delta n_\rr{e}\right) = 0
\end{equation}
in dimensionless units. This has the solution
\begin{equation}\label{elesol}
\delta n_\rr{e} = \varphi + C,
\end{equation}
where $C$ is constant along the magnetic field. Na\"ively, one could set this constant to zero and arrive at the HME. However, to correctly determine the value of $C$, one must use an additional constraint. 

Physically, electrons that evolve adiabatically will experience no radial transport, and so the density perturbation averaged over a flux surface must vanish. Thus, taking the flux-surface average of \eq{elesol} leads to $C = -\za{\varphi}_\psi$, or
\begin{equation} \label{properer}
\delta n_\rr{e} = \varphi - \za{\varphi}_\psi,
\end{equation} 
where $\za{\ldots}_\psi$ denotes the flux-surface average which, in two dimensions, is equivalent to the zonal average $\za{f}$,
\begin{equation}
\za{f} \doteq \frac{1}{L_y}\int_0^{L_y}\dd y\, f.
\end{equation}
Equation~\eq{poisson} is thus modified to now read
\begin{equation}\label{mHMp}
\zeta = \nabla_\perp^2\varphi - \varphi + \za{\varphi} \doteq (\nabla_\perp^2 - \h{\alpha})\varphi,
\end{equation}
where $\h{\alpha}$ is an operator that is zero when acting on zonal modes and is unity otherwise.  Notice that unlike~\eq{HMpol}, the adiabatic electron response given by~\eq{properer} \emph{does not} disappear in the Poisson bracket~\citep{jenko_2000}, so the $\bb{E}\btimes\bb{B}$ nonlinearity \emph{does} appear in the mHME, the importance of which is discussed in section~\ref{analytics}.

The Hasegawa-Mima equation that uses this form of the modified vorticity is referred to as the modified-Hasegawa-Mima equation (mHME). While the nonlinear interaction is changed by this new Poisson equation, it still conserves both the energy density $\mathcal{E}$ and the generalized enstrophy density $\mathcal{Z}$, where $\zeta$  in $\mathcal{Z}$ is now given by \eq{mHMp}. It is helpful to define the zonal ($k_y = 0$) and drift-wave ($k_y \ne 0$) energy densities,
\begin{align}
	\mathcal{E}^\rr{ZF} &\doteq\frac{1}{L_x} \int_0^{L_x}\dd x \, \frac{1}{2}\partial_x \za{\varphi}^2, 
\\	\mathcal{E}^\rr{DW} &\doteq\frac{1}{L_xL_y} \int_0^{L_x}\dd x \int_{0}^{L_y}\dd y \, \frac{1}{2}\left[(\nabla_\perp \varphi')^2 + \varphi'^2\right],
\end{align}
where $\varphi' = \varphi - \za{\varphi}$. These quantities will be helpful when determining whether a system is in a zonal-flow-dominated state or a turbulence-dominated state, and will be later used in Sec.~\ref{dns}. These definitions remain unchanged for both the Terry-Horton equation and modified-Terry-Horton equation (discussed below).

Both the HME and the mHME have been extensively studied in terms of both the modulational instability~\citep{Connaughton},  which concerns the instability of background drift wave to  a zonal flow, and the more general zonostrophic instability~\citep{Parker1,Parker2,Srinivasan1}, which encompasses the instability of \emph{any} statistically homogeneous steady state (including steady states of single realizations) to a zonal flow.

\subsection{The Terry-Horton Equation} \label{sTHE}

Another shortcoming of the HME is the lack of irreversibility. Studies of the HME usually add dissipation by hand and are stochastically forced to drive the system. An alternative to adding forcing is to include destabilizing effects in the electron response obtained iteratively from kinetic theory. These effects materialize most simply as an additional non-Hermitian operator $\THd$ in the electron response,
\begin{equation}
\delta n_\rr{e} = (1 - \ji \THd)\varphi .
\end{equation}
This results in a modification of the Poisson equation~\eq{poisson}, leading to
\begin{equation}
\zeta = (\nabla_\perp^2 + \ji\THd - 1)\varphi ,
\end{equation}
or
\begin{equation}
\h{\zeta} = -(k_\perp^2 - \ji \THd_\bb{k} + 1)\h{\varphi} 
\end{equation}
when Fourier transformed. Here, $\THd_\bb{k}$ is a real variable that depends on the wavevector $\bb{k}$.
%, and the Poisson wavenumber $\oo{\bb{k}}$ has been defined
 The resulting system is named the Terry-Horton Equation (THE) and has been extensively studied~\citep{TerryHorton1,TerryHorton2}. It has, however, fallen out of favour with respect to the more rigorously derived Hasegawa-Wakatani system, which has now become the standard system of equations for dealing with linearly unstable drift waves in a two-field fluid system.

 The non-adiabatic part of the electron response $\THd_\bb{k}$ for the THE can be chosen to describe various types of physical mechanisms. As an example, for the untrapped collisionless electron-wave resonance (the `universal' mode), one has in physical units~\citep{Tang78} 
\begin{equation}
\ji \THd_\bb{k} = \ji(\jpi/2)^{1/2} \left[\omega_\bb{k} - \omega_{*\rr{e}}\left(1 - \frac{1}{2}\eta_\rr{e}\right)\right] \Big/ |k_\parallel| v_\rr{the},
\end{equation}
where $\omega_{*\rr{e}} \doteq  c k_y T_\rr{e}/ e B_0 L_n$ is the electron diamagnetic drift frequency, $c$ is the speed of light, $B_0$ is the background magnetic field magnitude, and $\eta_\rr{e} \doteq \dd\, \ln T_\rr{e} / \dd\, \ln n_0$.
To proceed, $\omega_\bb{k}$ is taken to be the frequency of the linearized HME. Then, in dimensionless units,
\begin{equation} \label{deltak}
\ji \THd_\bb{k} = \ji\delta_0 k_y \left(\frac{k_\perp^2}{1 + k_\perp^2} - \frac{1}{2} \eta_\rr{e}\right),
\end{equation}
where $\delta_0 = (\jpi/2)^{1/2}(m_\rr{e}/m_\rr{i})^{1/2}/|k_\parallel L_n|$ is a constant of order unity that parameterizes the parallel wavenumber $k_\parallel$ (as $L_n$ is already parameterized by $\beta$).
One arrives at the original form of $\THd_\bb{k} = \delta_0 k_y(k_\perp^2 - \eta_\rr{e}/2)$ given by~\citet{TerryHorton1} by taking the long-wavelength ($k_\perp^2 \ll 1$) limit.
As another example, for trapped collisional electron dynamics, 
\begin{equation} \label{deltak2}
\ji\THd_\bb{k} =\ji\delta_0 k_y ,
\end{equation}
 where $\delta_0 = \eta_\rr{e}(\epsilon /2)^{1/2}(6/\jpi^{1/2})/\nu_\rr{eff}$ is a constant of order unity, $\epsilon$ is the inverse aspect ratio and $\nu_\rr{eff}$ is the collisional detrapping rate. A thorough review of such mechanisms is given by~\citet{Tang78}.

In addition to instability, the THE equation introduces non-zero particle transport. For instance, if one takes the non-adiabatic electron response to be simply $\ji\THd_\bb{k} = \ji k_y $, one obtains a nonzero particle flux
\begin{align}
\Gamma_n &\doteq \int \dd x \dd y \, v_x \delta n_\rr{i}
= \int \dd x \dd y \, \zeta \partial_y \varphi 
= \int \dd x \dd y \,(\partial_y \varphi)^2,
\end{align}
which is positive-definite. 
Another important feature of the THE is that it has only one nonlinearly conserved quantity, the generalized vorticity $\zeta$, which can alter the system's ability to experience a dual cascade. For instance, $\THd_\bb{k}$ as given by \eq{deltak2} becomes much larger relative to $\nabla_\perp^2$ at large scales, leading to degeneracy between $\mathcal{E}$ and $\mathcal{Z}$. This specific form of $\THd_\bb{k}$ has been shown by~\cite{Liang1993} to disable the dual cascade. However,  $\THd_\bb{k}$ as given by \eq{deltak} is small relative to $\nabla_\perp^2$ at both small and large scales. For this type of system the dual cascade is expected to remain prevalent.

\section{The Modified Terry-Horton Equation}\label{sMTHE}

\subsection{Description}
The model that is the focus of this article is a modified version of the THE that is designed to capture the essential zonal physics found in gyrokinetic simulations via a corrected electron response
\begin{equation}
\delta n_\rr{e} = (1 - \ji \THd)\varphi - \za{\varphi}_\psi.
\end{equation}
The resulting system is hence referred to as the modified Terry-Horton Equation (mTHE), and is given by
\begin{equation}\label{mTHE}
\frac{\partial \zeta}{\partial t} + \bb{v}\bdot \bb{\nabla} \zeta  = \beta \frac{\partial \varphi}{\partial y} - \h{\alpha} D\zeta,
\end{equation}
where $\bb{v} = (-\partial_y \varphi,\partial_x \varphi)$ and
\begin{equation}
\h{\zeta} = -(k_\perp^2 - \ji \THd_\bb{k} + \h{\alpha}_\bb{k})\h{\varphi}.
\end{equation}
 in Fourier space. 
Here, $D$ is a damping operator that, in Fourier space, is assumed to be even in both $k_x$ and $k_y$.
This model contains two modifications to the original THE. First, the adiabatic electron response is modified to ensure that electrons do not respond to a potential that is constant along a flux surface. This results in an enhancement of the zonal interaction between drift waves in the nonlinear term. The second modification is the appearance of the $\h{\alpha}$ operator in front of the damping term, which ensures that only the drift-wave modes ($k_y \ne 0$) are linearly damped. 
By doing so, the mTHE is made to model the residual Rosenbluth-Hinton zonal states~\citep{rosenbluth} witnessed in the simulations performed in~\citet{Dimits}. As a result, any state that consists purely of zonal flows is a steady-state solution to \eq{mTHE}. The damping on the drift waves can then be interpreted to be related to Landau damping of the potential fluctuation. As this is a fluid model, it is agnostic to the eventual fate of the fine-scale velocity-space structure that would result in the ion distribution function.  

One may also add a separate damping component to the zonal flows, as was done in~\citet{lin}. This damping typically results in bursty behaviour involving transitions between zonally dominated states and turbulence-dominated states within what would normally be the region of the Dimits shift. While this phenomenon is interesting in its own right, it is not touched upon in this article.

A question that should be raised is whether or not adding non-adiabatic effects to the electron response will also require modification to the constant of integration when solving the parallel electron force balance. However, electrons that are close to adiabatic should still not respond to a potential that is constant along a flux surface. For the remainder of this paper, I shall assume that these non-adiabatic effects are sufficiently small so as to not affect the adiabatic component of the response. Finally, I require that the flux-surface average of the non-adiabatic electron response $\za{\ji\THd \varphi}_\psi  = 0$, which is the case for the examples noted in Sec.~\ref{sTHE}. This is akin to saying that the background density gradients lie in the radial direction.

 \subsection{Linear Properties}
 
 The eigenvalues of \eq{mTHE} for the linearization around the zero state can be readily calculated. For drift-wave modes, the growth rates and frequencies for a time dependence of $\je^{\lambda_\bb{k} t}$ where $\lambda_\bb{k} = \gamma_\bb{k} - \ji \omega_\bb{k}$ are, respectively,
 \begin{align}
\gamma_\bb{k} &= -D_\bb{k} + \frac{\beta k_y \THd_\bb{k}}{(1 +k_\perp^2)^2 + \THd_\bb{k}^2},
 \\ \omega_\bb{k} &=  \frac{\beta k_y (1+k_\perp^2)}{(1+k_\perp^2)^2 + \THd_\bb{k}^2}.
 \end{align}
 For zonal modes, both are identically zero by construction, as I have neglected collisional damping.
 
 \subsection{Quasilinear Model}
 
 I also consider numerous approximations to \eq{mTHE}. The typical first step is to decompose the modified vorticity into zonal and non-zonal components, viz. $\zeta = \za{\zeta} + \zeta'$. By zonally averaging \eq{mTHE} and subtracting the result from the original equation, one obtains the new system of equations
 \begin{subequations}\label{QLset}
\begin{align}
\frac{\partial \zeta'}{\partial t} &= - U \frac{\partial \zeta'}{\partial y} - u' \frac{\partial^2 U}{\partial x^2}  + \beta \frac{\partial \varphi'}{\partial y}+ D \zeta' - F' \label{QLf},
\\ \frac{\partial U}{\partial t} &= -\frac{\partial}{\partial x}\za{u'v'} -  (\langle   u' \THd \varphi'\rangle- \ba{ u' \THd\varphi'}) \label{QLzf},
\end{align}
\end{subequations}
where $u' \doteq -\partial_y \varphi'$, $v' \doteq \partial_x \varphi'$,  $U(x) \doteq \partial_x \za{\varphi}$ is the zonal velocity, $F' \doteq\pb{\varphi'}{\zeta'} - \za{\pb{\varphi'}{\zeta'}} $, and  $\ba{\cdots}$ denotes the total spatial average. The last term in \eq{QLzf} results from integrating in $x$ to arrive at the equation for zonal velocity and is chosen to ensure mathematical consistency for $k_x =0$. 
So far, these equations are an exact description of the original equation (\ref{mTHE}). One arrives at the quasilinear system (QL) when the eddy-eddy interactions in the fluctuation equation (\ref{QLf}) are neglected, which is done by simply setting $F' = 0$. 

The physical effects that the quasilinear equations neglect are the eddy-eddy self-interactions. However, they do retain the eddy-eddy interactions that act on zonal modes, which appear in the form of a Reynolds stress (first term on the right-hand side of \eq{QLzf}), as well as a term unique to the Terry-Horton equation (second term on the right-hand side of \eq{QLzf}). The latter term describes radial $\bb{E}\btimes\bb{B}$ advection of the background electron gradient for the model $\THd_\bb{k}$ given by \eq{deltak}.
  It has already been shown that the quasilinear system captures the essential qualitative aspects of systems that are zonally dominant~\citep{Parker1,Farrell2}, and has provided many key insights into the role of zonal-flow-catalyzed energy transfer in the stabilization of drift-wave turbulence.

 \subsection{Four-Mode Truncation}\label{4mtsec}
 
Previous work on the HME and the two-field ITG model has concentrated on low-order Galerkin truncations, focusing on the modulational instability of a single mode to calculate zonal-flow growth rates, as well as on calculating the Dimits shift using the tools of dynamical systems~\citep{kolesnikov1,kolesnikov2}. To connect this article to this previous work, I formulate a four-mode truncation (4MT) of \eq{mTHE};  in Fourier space,
%\begin{widetext}
\begin{align}
\frac{\partial \varphi_\bb{k}}{\partial t} &=  (\gamma_\bb{k}-\ji\omega_\bb{k} )\varphi_\bb{k}+\frac{1}{\h{\alpha}_\bb{k} -  \ji \THd_\bb{k} + k_\perp^2} \sum_{\bb{k}_1,\bb{k}_2}k_{1x}k_{2y}   \varphi_{\bb{k}_1}\varphi_{\bb{k}_2} 
 \nonumber
\\ & \hspace{2cm} \times \delta_{\bb{k},  \bb{k}_1 + \bb{k}_2}[\h{\alpha}_{\bb{k}_2} - \h{\alpha}_{\bb{k}_1} - \ji(\THd_{\bb{k}_2} -\THd_{\bb{k}_1})+ k_{2\perp}^2 - k_{1\perp}^2],
\end{align}
where $\delta_{\bb{a},\bb{b}}$ denotes the Kronecker delta.
In the 4MT, I retain a pure drift-wave mode $\bb{p} = (0,p_y)$, a pure zonal mode $\bb{q} = (q_x,0)$, and two sidebands $\bb{r}_\pm = (\pm q_x, p_y)$. In addition, the complex conjugate modes are retained in order to satisfy the reality condition. I also assume that $\THd_{\bb{r}_-} = \THd_{\bb{r}_+} \doteq \THd_{\bb{r}}$, which is the case for the two examples of $\THd_\bb{k}$ given in Sec.~\ref{sTHE}. Finally, $\h{\alpha}_\bb{q}$ is kept arbitrary to compare with previous results. 

The resulting set of equations is
\begin{subequations} \label{4mt}
\begin{align}
\partial_t \varphi_\bb{p} &=(\gamma_\bb{p} -\ji\omega_\bb{p})\varphi_\bb{p}  + M_\bb{p}(\varphi_{\bb{r}_-}\varphi_{\bb{q}}-\varphi_{\bb{r}_+}\varphi^*_{\bb{q}}), \label{4mt1}
\\ \partial_t \varphi_{\bb{r}_+} &=  (\gamma_\bb{r} - \ji\omega_\bb{r}) \varphi_{\bb{r}_+} +M_\bb{r}\varphi_{\bb{p}}\varphi_{\bb{q}},  \label{4mt2}
\\  \partial_t \varphi_{\bb{r}_-}^* &= (\gamma_\bb{r} + \ji\omega_\bb{r})\varphi_{\bb{r}_-}^* - M_\bb{r}^*\varphi_{\bb{p}}^*\varphi_{\bb{q}}, \label{4mt3}
\\  \partial_t \varphi_\bb{q} &= \frac{q_x p_y}{\h{\alpha}_\bb{q} + q_x^2}[ q_x^2(\varphi_{\bb{r}_+}\varphi_{\bb{p}}^* - \varphi_{\bb{r}_-}^* \varphi_{\bb{p}}) - \ji \THd_+(\varphi_{\bb{r}_+}\varphi_{\bb{p}}^* + \varphi_{\bb{r}_-}^* \varphi_{\bb{p}})], \label{4mt4}
\end{align}
\end{subequations}
along with their complex-conjugate counterparts. Here,  $\THd_\pm \doteq \THd_\bb{p} \pm \THd_\bb{r}$, and the $M_\bb{k}$'s are mode-coupling coefficients of the interaction between zonal flows and drift waves and are given by
 \begin{subequations} \label{mcc}
\begin{align}
M_\bb{p} &= q_x p_y \frac{1 + p_y^2 - \ji \THd_\bb{r} - \h{\alpha}_\bb{q} }{1 + p_y^2 - \ji \THd_\bb{p}},
\\ M_\bb{r} &= q_x p_y \frac{1 - q_x^2 + p_y^2 - \ji \THd_\bb{p} - \h{\alpha}_\bb{q} }{1 + q_x^2 + p_y^2 - \ji \THd_\bb{r}}.
\end{align}
\end{subequations}
It will be useful to express these coefficients in terms of their real and imaginary parts,
 \begin{subequations}
\begin{align}
	M^\rr{Re}_\bb{p}&= \frac{q_x p_y[(1 + p_y^2)^2  +\THd_\bb{p}\THd_\bb{r} - \h{\alpha}_\bb{q}(1 + p_y^2)]}{(1 + p_y^2)^2 + \THd_\bb{p}^2},
\\	M^\rr{Im}_\bb{p}&= \frac{q_xp_y[ \THd_-(1+p_y^2) - \h{\alpha}_\bb{q}\THd_\bb{p}] }{(1+ p_y^2)^2 + \THd_\bb{p}^2},
\\	M^\rr{Re}_\bb{r}&= \frac{q_x p_y[(1 + p_y^2)^2 - q_x^4 +\THd_\bb{p}\THd_\bb{r} - \h{\alpha}_\bb{q}(1+q_x^2 + p_y^2)]}{(1+q_x^2 + p_y^2)^2 + \THd_\bb{r}^2},
\\	M^\rr{Im}_\bb{r}&= -\frac{q_xp_y[\THd_+q_x^2 + \THd_-(1+p_y^2) + \h{\alpha}_\bb{q}\THd_\bb{r}] }{(1+q_x^2 + p_y^2)^2 + \THd_\bb{r}^2}.
\end{align}
\end{subequations}
 To obtain these, I have used the fact that $\THd_\bb{-k} = -\THd_\bb{k}$.

One quantity that can be readily calculated for the 4MT is the point at which the system loses its ability to saturate (becomes globally unstable). For an $N$-mode truncation, this is done by considering the phase space of modes consisting of the $2N$ real and imaginary components of the mode amplitudes and their corresponding velocities that define the equations of motion. \citet{TerryHorton2} have shown that the necessary condition for the system to be absolutely stable is for the phase space at any arbitrary point to be volume contracting. Equivalently, the sum of the growth rates of every mode must be lesser or equal to zero ($\sum_\bb{k}\gamma_\bb{k} \le 0$). This condition provides insight into a fundamental property of diagonalized partial differential equations with a Poisson-bracket-type quadratic nonlinearity and constant linear coefficients: when one linearizes around \emph{any} arbitrary point in the phase-space of modes, the diagonal of the resulting matrix can only be populated  by linear terms. Thus, if the trace of that matrix is positive, then there exists at least one positive eigenvalue and the system is globally unstable. While this trace is usually negative once a sufficient amount of stable modes is included (e.g. modes at small scales that are viscously damped), this has interesting consequences when a severe truncation of the system can be justified. This will be seen in section~\ref{DScalc}.

For the 4MT considered here, the necessary condition for absolute stability is 
\begin{equation}\label{phasespace}
\gamma_\bb{p} + 2 \gamma_\bb{r} \le 0,
\end{equation}
 though nonlinear mode coupling tends to approximately make this the sufficient condition as well, in the sense that only linear gradients near the necessary threshold seem to destabilize the entire system. This is seen in numerical simulations of the 4MT (see figure~\ref{fluxes}).

\section{Numerical results} \label{numerics}

\begin{figure}
\centering
\includegraphics[scale=0.75]{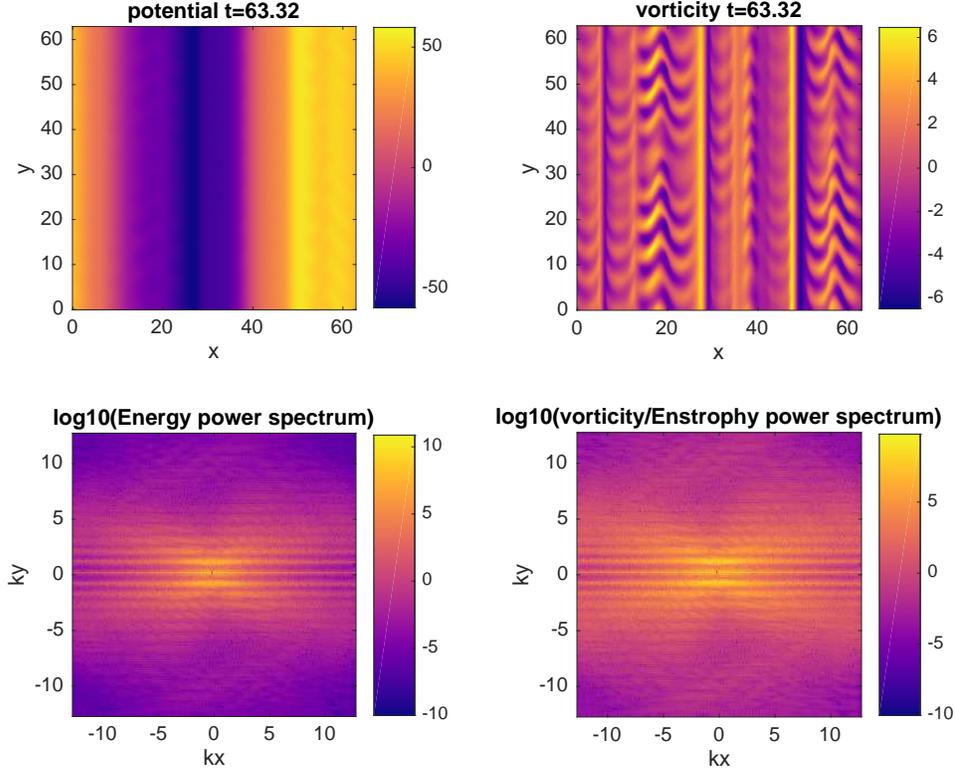}
\caption{\label{snapshot}Snapshot of the evolution of the NL2 from direct numerical simulation with $D  = 1 - 0.01 \nabla_\perp^2$, $\beta = 5.5$ and $\THd_\bb{k} = 1.5 \ji k_y$. This value of $\beta$ corresponds to slightly below the end of the Dimits shift. Top left displays the evolution of the potential $\varphi$, while the top right displays the evolution of the modified vorticity $\zeta$, with their respective power spectra displayed underneath.
%Clicking on the figure shows the simulation running from $t = 0$ to $t = 500$, capturing the full evolution of the Dimits shift.
}
\end{figure}

\subsection{Setup}

The nonlinear (NL) system (\ref{mTHE}) and quasilinear (QL) system (\ref{QLset})  are simulated pseudospectrally and dealiased on a square Cartesian grid with $L = 20 \jpi$ and $N = 256$ on each side. Time stepping is performed using third-order Adams-Bashforth with an integrating factor. The vorticity at $t  = 0$ is initialized with Gaussian noise of zero mean and standard deviation $5 \times 10^{-3}$, with no energy in zonal modes. The random number generator used for the initial state is initialized with the same seed for all simulations. 
For all simulations in this section, $D = \mu - \nu \nabla_\perp^2$ where $\mu = 1$ and $\nu = 10^{-2}$. Here, friction $\mu$ is added to dissipate energy at large-scales and is added as a preventative measure against a possible inverse cascade. (It is shown in Sec.~\ref{analytics} that the qualitative aspects of the simulations do not depend on the choice of damping operator.)
To simulate the four-mode truncation, I choose modes with $q_x = p_y = 1$ (the dependence of this system on the specific values of $q_x$ and $p_y$, as well as the damping operator $D$, is discussed in Sec.~\ref{analytics}). Finally, the value of $\beta$ at which the Dimits shift ends is denoted as a critical density gradient $\beta^*$. The size of the shift then is $\Delta \beta \doteq \beta^* - \beta_\rr{lin}$.

I run separate simulations using the different  $\THd_\bb{k}$'s given by \eq{deltak}  and \eq{deltak2}, the former with  $\delta_0 = 2$  and $\eta_\rr{e} = 0$, and the latter with $\delta_0 = 1.5$. To differentiate between the two systems, I denote the former by (NL1), (QL1), etc., and the latter by (NL2), (QL2), etc. For results that are insensitive to the details of $\THd_\bb{k}$, I simply use the unnumbered abbreviations. The parameters given above correspond to a threshold of linear instability at $\beta_\rr{lin} \approx 4.74$ for $\THd_\bb{k}$ given by \eq{deltak}, and $\beta_\rr{lin} \approx 4.21$ for $\THd_\bb{k}$ given by \eq{deltak2}.

\subsection{Direct Numerical Simulation}\label{dns}

Figure~\ref{snapshot} contains an animation that shows the evolution of the NL2 system~(\ref{mTHE}) with $\THd_\bb{k} = 1.5 \ji k_y$ within the Dimits-shift regime ($\beta = 5.5$). Additional movies with varying values of $\beta$ for the nonlinear system ($\beta = 4.5$, $6.5$) as well as a movie of the quasilinear system~(\ref{QLset}) with $\beta = 5.5$ are included as supplementary material, which can be accessed online at \texttt{https://arxiv.org/abs/1704.05406}.
 Every simulation begins with a short period of linear damping of stable drift-wave modes and linear growth of unstable ones. If the system begins below the threshold of linear instability ($\beta < \beta_\rr{lin}$) then all modes damp and the final state is the zero state; otherwise, growth of the drift-wave modes are clustered around the most unstable mode with growth rate $\gamma = \gamma_\rr{max}$.
 Because the perturbation level at this stage is quite small ($\varphi_\bb{k}\ll 1$), these modes are allowed to grow without any influence from the nonlinear interaction. As the drift-wave energy density $\mathcal{E}^\rr{DW}$ grows and becomes of order unity, the nonlinear interaction becomes effective in transferring energy to zonal modes, resulting in fast growth in zonal energy.

\begin{figure}
\centering
\includegraphics[scale=0.90]{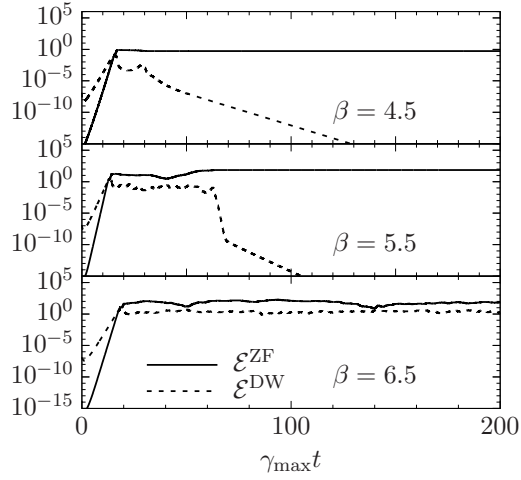}
\caption{\label{DNScases} Direct numerical simulation of the NL2 system (\ref{mTHE}) with $D  = 1 - 0.01 \nabla_\perp^2$ and $\THd_\bb{k} =1.5\ji k_y$ for three values of density-gradient parameter $\beta$. The solid line denotes the zonal component of the energy density, while the dashed line denotes the energy density in all non-zonal modes.}
\end{figure}

Once $\mathcal{E}^\rr{ZF} \sim \mathcal{E}^\rr{DW}$, the nonlinear interaction becomes important and two different scenarios can take place. One possibility is for the system to find a stable zonal state that occurs after a burst of turbulent interactions~\citep{kolesnikov1,kolesnikov2}. Once a stable zonal spectrum is found, the system then relaxes to a pure zonal state with the non-zonal modes damped away to zero. For simulations near the end of the Dimits shift, the system can cycle through a number of zonal spectra until finding one that is ultimately stable. The time for this to happen typically increases with $\beta$, though for a given realization this may not always strictly be true.
The other possibility is that no such stable zonal spectrum can be reached, if it even exists. 
In this case the system remains in a turbulent state with finite particle flux. As an example, the nonlinear system with $\beta = 6.5$ and $\THd_\bb{k} =\ji\delta_0 k_y$ was run for a time of $t = 25\,000$ without ever reaching a stable zonal state. The system with $\THd_\bb{k}$ given by \eq{deltak}, on the other hand, tends to always find a stable zonal state, thus it does not exhibit a finite Dimits shift. This situation is further discussed in the last paragraph of this section.
 The qualitative behaviour of both the QL system and 4MT are also similar, though the quantitative aspects of the 4MT, which will be discussed in section~\ref{tertiary}, are quite different.

Figure~\ref{DNScases} shows the evolution of both zonal and drift-wave energy densities of the NL2 system for three values of $\beta$ versus time scaled by the respective growth rate of the fastest growing mode. The first case ($\beta = 4.5$) is slightly above the threshold for linear instability. Here, the stable zonal state is found in a short time, as can be seen in the supplemental movie \texttt{DS\_b4.5\_NL.mp4}.  The second case is near the end of the Dimits shift ($\beta = 5.5$).  Now the system spends significantly more time cycling through several zonal spectra until a final one is found. This is shown in figure~\ref{snapshot}. Finally, the third case is past the end of the Dimits shift ($\beta=6.5$). For this case, turbulence persists and no stable zonal state is found (see supplemental movie \texttt{DS\_b6.5\_NL.mp4}). These are quantitatively similar for the quasilinear system, and a representative evolution of this system is shown in the supplemental movie \texttt{DS\_b5.5\_QL.mp4}.

\begin{figure}
\centering
\includegraphics[scale=0.90]{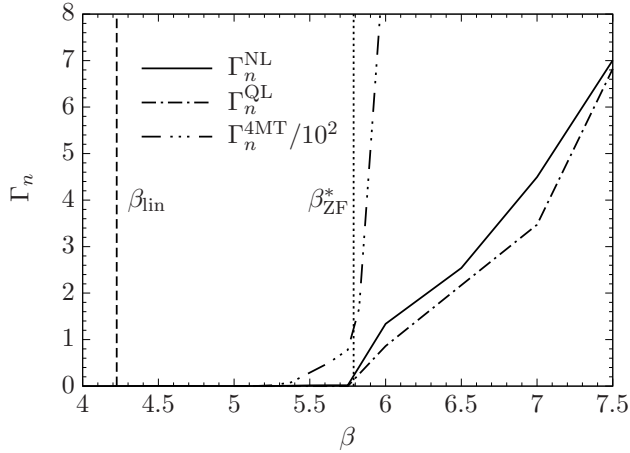}
\caption{\label{fluxes}Particle flux for the NL2, QL2 and 4MT2 systems with $D  = 1 - 0.01 \nabla_\perp^2$ and  $\THd_\bb{k} = 1.5 \ji  k_y$ as a function of ion-density-gradient parameter $\beta$. The medium-dashed line denotes the linear threshold for instability $\beta_\rr{lin}$. The dotted-line line denotes predicted end of the Dimits shift $\beta^*$ given from the solution of \eq{thething}.
}
\end{figure}

Figure~\ref{fluxes} shows the long-time turbulent flux as a function of $\beta$ for the NL2, QL2 and 4MT2 systems with $\THd_\bb{k} = 1.5 \ji  k_y$. (This plot serves the same purpose as figure 3 in~\citealt{Dimits}.) As expected, the Dimits shift is observed as a distinct lack of any flux immediately beyond the point of linear instability. Once the end of the Dimits shift is reached ($\beta^* \approx 5.75$ for both the NL2 and QL2 systems), turbulence is again allowed to persist and turbulent transport ensues. 
It is important to note that the Dimits shift is quantitatively similar between the NL and the QL systems. This is a rather profound result, as the NL system contains additional avenues of energy dissipation through the direct cascade, where turbulent eddies continuously self-interact, forming smaller turbulent eddies which eventually leads to damping. On the other hand, eddies in the QL system can only be sheared via zonal flows. However, this shearing is the dominant mechanism when the enhancement of the nonlinear zonal interaction is introduced via a physical adiabatic electron response. (I shall show what this means in Sec.~\ref{tertiary}.) As a result, energy transfer is principally in the direction of $k_x$.
This has interesting consequences, the important one being that a QL-like closure, such as CE2, suffices to capture all the relevant physics needed for the Dimits shift. 

It is also important to note the discrepancy of the size of the Dimits shift, as well as the saturation levels, of the 4MT.  This discrepancy is simply a result of the fact  that there is only a single stable mode that can accept energy. This also explains the sharp discrepancy between the saturated levels of flux between the NL/QL and 4MT; because there are far fewer stable modes, the effective damping rate is greatly reduced in the 4MT, resulting in larger levels of saturation when the system strikes a balance between energy production (which has not changed significantly for the 4MT as the most unstable mode is retained) and energy dissipation (which has). Finally, the necessary condition for absolute stability is given by~\eq{phasespace}. This gives a gradient threshold for saturation of $\beta^\rr{sat} \approx 6.0$ for the parameters given in figure~\ref{fluxes}. This agrees well with what is observed numerically.

\begin{figure}
\centering
\includegraphics[scale=0.90]{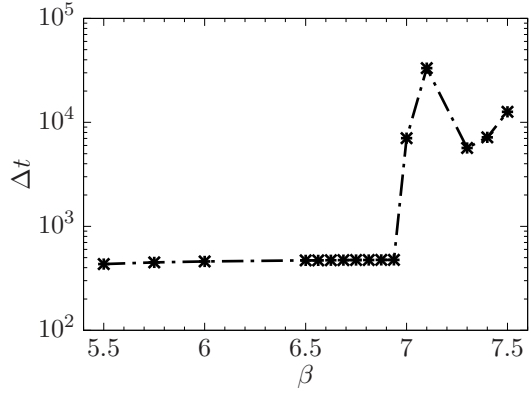}
\caption{\label{time} The time taken to reach a steady zonal state $\Delta t$ for the NL system 
with $D  = 1 - 0.01 \nabla_\perp^2$ and  $\THd_\bb{k}$ given by \eq{deltak} with $\delta_0 = 2$ and $\eta_\rr{e} = 0$ as a function of ion-density-gradient parameter $\beta$. Two separate regimes can clearly be seen, one where the system quickly settles into a zonal state  ($\beta < 7$), and one where the system takes an extended amount of time to settle ($\beta > 7$). }
\end{figure}

I have mentioned that the system with $\THd_\bb{k}$ given by \eq{deltak} does not exhibit a finite Dimits shift; rather, a steady zonal state seems to be always found. Even so, two different regimes materialize, as shown in figure~\ref{time}. Here, the  time taken to reach a steady zonal state is defined as $\Delta t \doteq t_\rr{f} - t_\rr{i}$, where $t_\rr{i}$ is the first time where  $\mathcal{E}^\rr{ZF} = \mathcal{E}^\rr{DW}$, and $t_\rr{f}$ is the first time where $\mathcal{E}^\rr{DW}/\mathcal{E}^\rr{ZF} = 10^{-6}$.
In the first regime where $\beta <7$, steady zonal states are found within $\Delta t \sim 500$. When $\beta > 7$, this time increases by at least a full order of magnitude. One possible explanation is that in the first regime, the zonal shearing associated with the Dimits shift is operational and zonal states that can sufficiently quench the drift wave turbulence are quickly found,  while in the second regime some other channel of energy transfer is eventually established, leading to stable solutions. A potential candidate for energy transfer in this second regime could be the nonlocal cascade of enstrophy to small scales that is typically associated with the dual cascade. More work needs to be done to better understand this regime; that is left for future study.

\section{Analytical Results} \label{analytics}

I begin this section with a linear stability analysis of both the secondary and tertiary instabilities for the 4MT. With these results in hand, the size of the Dimits shift can be estimated for the fully nonlinear system, and the underlying mechanism behind the shift in this system can be well understood in terms of efficient energy transfer between unstable and stable modes.

\subsection{Zonal growth rates of the secondary instability}\label{zGrowth}

At the beginning of the simulation, the most unstable drift-wave mode grows without any influence from the zonal or side-band modes in what I shall call the secondary regime. One can then solve~(\ref{4mt}b--d) with $\varphi_\bb{p} = \varphi_0 \je^{(\gamma_\bb{p} + \ji \omega_\bb{p})t}$, which renders the remaining equations a linear system:
\begin{subequations}\label{4mtl}
\begin{align}
\label{4mtl1} \partial_t \varphi_{\bb{r}_+} &=  (\gamma_\bb{r}  - \ji \omega_\bb{r}) \varphi_{\bb{r}_+} +M_\bb{r}\varphi_0 \je^{\gamma_\bb{p} t - \ji \omega_\bb{p} t }\varphi_{\bb{q}},
\\ \label{4mtl2}\partial_t \varphi_{\bb{r}_-}^* &= (\gamma_\bb{r} + \ji\omega_\bb{r})  \varphi_{\bb{r}_-}^* -M_\bb{r}^*\varphi_0^* \je^{\gamma_\bb{p} t + \ji \omega_\bb{p} t }\varphi_{\bb{q}},
\\ \partial_t \varphi_\bb{q} &= \frac{q_xp_y\je^{\gamma_\bb{p} t}}{\h{\alpha}_\bb{q} + q_x^2}[q_x^2 ( \varphi_{\bb{r}_+}\varphi_0^*\je^{\ji\omega_\bb{p} t} -  \varphi_{\bb{r}_-}^* \varphi_0 \je^{-\ji\omega_\bb{p} t}) \nonumber
\\  \label{4mtl3} & \hspace{2.5cm}- \ji \THd_+(\varphi_{\bb{r}_+}\varphi_0^* \je^{\ji\omega_\bb{p} t} + \varphi_{\bb{r}_-}^* \varphi_0 \je^{-\ji\omega_\bb{p} t})].
\end{align}
\end{subequations}
The goal is now to determine the growth rate of the zonal flow against this growing drift-wave mode. 

At this point, one can make a further approximation by focusing on the region where the zonal mode grows much faster than the drift-wave mode,  making the assumption $\gamma_\bb{p} = \gamma_\bb{r} = 0$. Alternatively one can also proceed, without any further assumptions, to derive an evolution equation for $\varphi_\bb{q}$; this is done here.

 To do so, the transformations 
$\varphi_{\bb{r}_+} = \varphi_0\varphi'_{\bb{r}_+} \je^{-\ji\omega_\bb{p} t}$ and  $\varphi^*_{\bb{r}_-} = \varphi_0^*\varphi'^*_{\bb{r}_-} \je^{\ji\omega_\bb{p} t} $ are made to eliminate the rapid oscillatory behaviour of the drift-wave mode. New equations of motion are then formulated for the variables $\Lambda_\pm \doteq \varphi'_{\bb{r}_+} \pm \varphi'^*_{\bb{r}_-}$. Defining  $\gamma_\pm \doteq \gamma_\bb{p} \pm \gamma_\bb{r}$ and $\omega_\pm \doteq \omega_\bb{p} \pm \omega_\bb{r}$,~(\ref{4mtl}a--c) become
\begin{subequations}\label{almost}
\begin{align}
	\partial_t \Lambda_+ &= \gamma_\bb{r}\Lambda_+ - \ji  \omega_-\Lambda_- + 2 \ji M_\bb{r}^\rr{Im}\je^{\gamma_\bb{p}t}\varphi_\bb{q},
	\\	\partial_t \Lambda_- &= \gamma_\bb{r}\Lambda_- - \ji  \omega_- \Lambda_+ + 2  M_\bb{r}^\rr{Re} \je^{\gamma_\bb{p}t}\varphi_\bb{q},
\\	\partial_t \varphi_\bb{q} &= \frac{|\varphi_0|^2 q_x p_y \je^{\gamma_\bb{p}t}}{\h{\alpha}_\bb{q} + q_x^2}\left(q_x^2\Lambda_- -  \ji \THd_+ \Lambda_+ \right).
\end{align}
\end{subequations}
 From here, it is a simple exercise to derive an ordinary differential equation for the zonal mode by combining (\ref{almost}a--c):
\begin{align}\label{ZFg}
\varphi'''_\bb{q} - A \varphi''_\bb{q} + \left(B - C \je^{2 \gamma_\bb{p}t}\right)\varphi'_\bb{q} - D \je^{2\gamma_\bb{p}t}\varphi_\bb{q} = 0,
\end{align}
where
\begin{subequations}
\begin{align}
A &\doteq 2\gamma_+, 
\\ B &\doteq  \omega_-^2 + \gamma_+^2,
\\ C &\doteq \frac{2 |\varphi_0|^2 q_x p_y}{\h{\alpha}_\bb{q}+q_x^2} \left(q_x^2 M^\rr{Re}_\bb{r} + \THd_+ M^\rr{Im}_\bb{r} \right),
 \\ D &\doteq \frac{2 |\varphi_0|^2 q_x p_y}{\h{\alpha}_\bb{q}+q_x^2} \Big[\gamma_-\left(q_x^2  M^\rr{Re}_\bb{r} +\THd_+M^\rr{Im}_\bb{r} \right) +  \omega_-\left(q_x^2 M^\rr{Im}_\bb{r} -\THd_+ M^\rr{Re}_\bb{r} \right)\Big].
\end{align}
\end{subequations}
The usual dispersion relation of the modulational instability for the HME is recovered by setting $\THd_\bb{k} = 0$ and $\h{\alpha}_\bb{q} = 1$~\citep{Connaughton}.

The goal is to analyze \eq{ZFg} in the asymptotic limits $t \rightarrow 0$ and  $t \rightarrow \infty$. The latter, however, is made difficult by the fact that $C$, $D \propto |\varphi_0^2 | \lll 1$. Futhermore,  $C$ and $D$ can have values with quite different magnitudes \emph{relative to each other}, exacerbating the situation. The limit $t \rightarrow \infty$ cannot be taken at face value then, as this asymptotic time may occur well beyond the range of validity of this approximation (that is to say, beyond the secondary regime).
 Thus, \eq{ZFg} must be analyzed in the asymptotic limit $t\rightarrow ``\infty"$, which is coarse-grained according to the relative size of the coefficients $A$, $B$, $C$ and $D$, based on a  given set of parameters $q_x$, $p_y$ and $\beta$. 

For $t \rightarrow 0$,  the $C$ and $D$ coefficients become subdominant. 
 The resulting equation is solved readily with solutions
\begin{equation}
y \sim \exp \left[ (\gamma_+ \pm \ji \omega_-)t\right]
\end{equation}
and  $y \sim c$, where $c$ is a constant. This corresponds to the case where the sideband mode also grows independently with growth rate $\gamma_\bb{r}$. The zonal flow growth rate is then $\gamma_+$, which is maximized at the largest scale of the box. This early type of growth is seen in both the simulations presented here (see figure~\ref{snapshot}) as well as the simulations of~\citet{rogers}.

\begin{figure}
\centering
\includegraphics[scale=0.90]{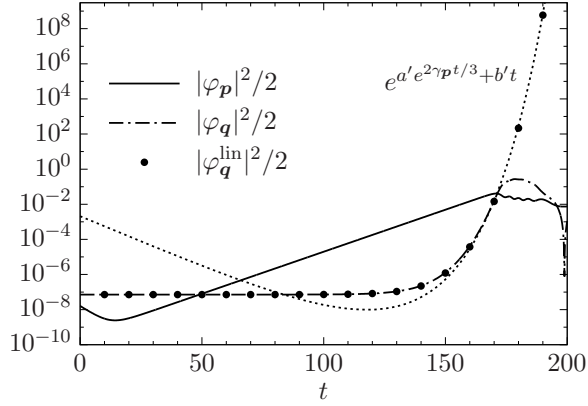}
\caption{\label{4MT_test}Typical energy evolution for the 4MT system as described in Sec.~\ref{zGrowth}  with $q_x = 0.6$, $p_y= 1.3$, $D = 1 - 0.01 \nabla_\perp^2$,  $\beta = 5$ and $\THd_\bb{k}$ given by \eq{deltak} with $\delta_0= 2$ and $\eta_\rr{e} =0$. The solid line denotes the drift-wave mode energy while the dot-dashed line denotes the zonal mode energy. The numerical solution to \eq{ZFg} is plotted with points, with the $t\rightarrow ``\infty"$ asymptotic behaviour being shown as a dotted line.}
\end{figure}

To analyze the limit $t \rightarrow ``\infty "$, the ansatz $\varphi_\bb{q} \sim e^{S(t)}$  is used in \eq{ZFg}, resulting in 
\begin{equation}
S''' +  3S'' S' + S'^3 - A (S'' + S'^2) + (B - e^{2\gamma_\bb{p}t}C)S' - e^{2 \gamma_\bb{p}t}D = 0.
\end{equation}
Two cases are now considered. First, let both $C$ and $D$ be of the same order. 
One then finds the leading order behaviours $\varphi_\bb{q} \sim \exp (- Dt / C)$ and
\begin{align} \label{sc1}
\varphi_\bb{q} \sim \exp (\pm a \je^{\gamma_\bb{p}t} + b t),
\end{align}
where
\begin{subequations}
\begin{align}
	a &\doteq \sqrt{C} / \gamma_\bb{p},
\\	b &\doteq  (AC + D - 3 \gamma_\bb{b} C)/ 2 C.
\end{align}
\end{subequations}
These apply equally well to the case where $D \rightarrow 0$. For the other case where $C$ is subdominant, one finds the leading order behaviour
\begin{align}\label{sc2}
\varphi_\bb{q} \sim \exp(I_3 a' \je^{2\gamma_\bb{p}t/3} + b' t),
\end{align}
where
\begin{subequations}
\begin{align}
	a' &\doteq\frac{3 D^{1/3}}{2 \gamma_\bb{p}},
\\	b' &\doteq  (A  - 2 \gamma_\bb{b} )/ 3,
\end{align}
\end{subequations}
and $I_3$ is a cube root of unity.  Both \eq{sc1} and \eq{sc2} represent instabilities of Kelvin-Helmholtz type ($\gamma \sim |q_x p_y \varphi_\bb{p} |^{\varepsilon} 
\sim |q_x v_x |^{\varepsilon}$ where $\varepsilon$ is a constant of order unity). This form of the growth rate agrees with the usual picture of the secondary instability given by \citet{rogers}, and is roughly maximized at scales between $k_y^{-1}$ and the sound radius (i.e. $k_x \sim  k_y $ to $k_x \rho_\rr{s}  \sim 1$).

 Again, it is emphasized that which asymptotic behaviour is relevant depends on the specific values of $q_x$, $p_y$, and $\beta$, as well as where the secondary regime ends. For instance, with $q_x = 0.6$, $p_y = 1.3$ and $\beta = 5$, $C$ is subdominant for $t \lesssim 200$, and so the second situation applies. Figure~\ref{4MT_test} shows the evolution of the drift-wave energy and the zonal energy for the 4MT with $\beta = 5$. Both the numerical solution of \eq{ZFg} and the scaling given by \eq{sc2} are displayed, showing excellent agreement.

Unfortunately, it is difficult to extend this analysis to the NL or QL systems in order to determine the dominant zonal mode, as the cumulative effect of the asymptotic behaviour of every mode is somewhat unclear. Even then, such a prediction would only predict a dominant zonal mode \emph{during the initial secondary stage}. There is no \emph{a priori} reason why this mode would remain the dominant one once the fully nonlinear interaction stage comes to an end.  More sophisticated approaches, such as the wave-kinetic equation~\citep{diamond2005,parker_2016,ruiz2017}, have been used in the past to calculate zonal growth rates with some success. However, this approach assumes a homogeneous background of drift-wave turbulence~\citep{krommes_kim}, which is not the case in this regime, and so the usefulness of the wave-kinetic equation for calculating zonal growth rates of the secondary instability within the Dimits shift regime is at this point uncertain.

\begin{figure}
\centering
\includegraphics[scale=1.0]{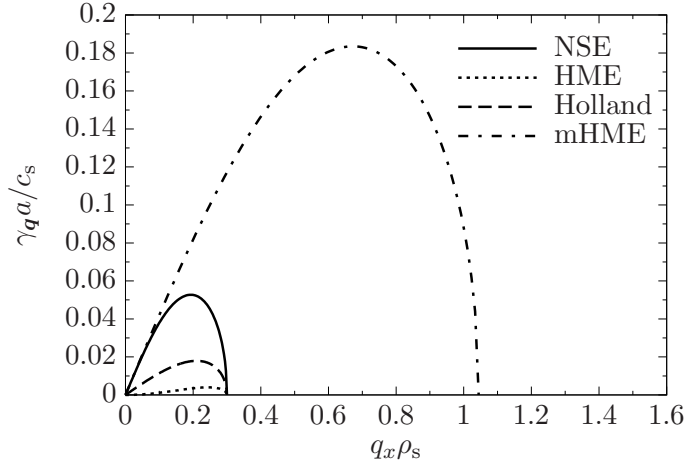}
\caption{\label{ZGR}  Zonal growth rate as a function of the zonal-flow radial wavenumber $q_x$ over a large-amplitude drift-wave background with poloidal wavenumber $p_y = 0.3$ and $\varphi_0 = 1$. The solid line denotes the growth rate using a standard two-dimensional incompressible Navier-Stokes equation (NSE, i.e. $\h{\alpha}$ is identically zero). The dotted line denotes the growth rate for the HME, while the dashed-dotted line denotes the growth rate for the mHME, both with $\beta = 0$. The dashed line denotes the mHME when the electron response is modified only within the partial time derivative, as studied by~\citet{Holland}. (From G.~\citealt{hammettLS}, private communication.)}
\end{figure}

In order to appreciate the effect of the physical adiabatic electron response given by~\eq{properer}, it is instructive to reconsider the above analysis in the limit of no growth or dispersion, retaining only the nonlinear interactions~(G.~\citealt{hammettLS}, private communication). This is akin to taking the strong turbulence limit. In addition, the Terry-Horton term $\THd$ is set to zero. Then~\eq{4mtl} immediately leads to the zonal flow growth rate 
\begin{equation}
\gamma_\bb{q} = |q_x p_y \varphi_0|\sqrt{2} \left(\frac{q_x^2}{\h{\alpha}_\bb{q} + q_x^2}\frac{ \h{\alpha}_\bb{p} - \h{\alpha}_\bb{q}  - q_x^2 + p_y^2  }{ \h{\alpha}_\bb{r}  + q_x^2 + p_y^2 }\right)^{1/2},
\end{equation}
where the $\h{\alpha}_\bb{k}$ operator has been kept arbitrary. 
This growth rate is plotted in figure~\ref{ZGR} using the HME and mHME with $p_y = 0.3$ and $\varphi_0 = 1$. In addition, the same growth rate using the two-dimensional incompressible Navier-Stokes equation (NSE) is calculated by  taking $\h{\alpha}_\bb{k} = 0$ identically. For $q_x$, $p_y \ll 1$, physical adiabatic electron response under the partial time derivative of~\eq{HME} leads to an enhancement of zonal flow growth by a factor of $1/(q_x p_y)$. A similar level of growth is also seen in the NSE. It is this correction to the partial time derivative that led to the idea that zonal flows would grow at the largest scales, forming the idea of what was called the `mesoscale' between microturbulence and equilibrium scales~\citep{diamond2005}. However, when the physical adiabatic electron response is taken into account \emph{in the Poisson bracket} (resulting in an $\bb{E}\btimes\bb{B}$ nonlinearity), the range of zonal flow scales in which energy is deposited is greatly increased, down to the scale given by
\begin{equation}\label{cutoff}
q_x^2 = 1 + p_y^2.
\end{equation} 
In tokamak plasmas, $p_y \ll 1$, and so this results in a \emph{nonlocal} transfer of energy from the primary drift wave to zonal flows at scales of order $\sim \rho_\rr{s}$, which is at odds with the conventional wisdom that energy flows to zonal flows at the \emph{largest} scales~\citep{diamond2005}. This was previously seen in both \citet{rogers} and \citet{guzdar}. When physical electron response is inconsistently taken into account under the partial time derivative but not in the Poisson bracket (as was done in~\citealt{Holland}), the zonal flow growth rate is greatly reduced. This is denoted in figure~\ref{ZGR} by a dashed line. While the above analysis of the zonal flow growth rate is based on a simplified four-mode truncation, the cutoff given by~\eq{cutoff} is also observed in statistical closures of the mHME, including CE2 and a geometrical-optics reduction of CE2~\citep{parker_2016}.
The importance of nonlocal energy transfer effected by the $\bb{E}\btimes\bb{B}$ nonlinearity is further elaborated upon in the next section.

\subsection{Zonal mode stability analysis of the tertiary instability} \label{tertiary}

The stability analysis of the previous section can be repeated to study the stability of a zonal mode to a drift-wave perturbation, otherwise known as the tertiary instability. For a background zonal state $\varphi_\bb{q} = \varphi_0$ of the 4MT, the linearized equations of motion are
 \begin{subequations}
\begin{align}
\partial_t \varphi_\bb{p} &=\lambda_\bb{p}\varphi_\bb{p} + M_\bb{p} (\varphi_0\varphi_{\bb{r}_-}-\varphi_0^*\varphi_{\bb{r}_+}),
\\ \partial_t \varphi_{\bb{r}_+} &= \lambda_\bb{r} \varphi_{\bb{r}_+} +M_\bb{r}\varphi_{\bb{p}}\varphi_0,
\\ \partial_t \varphi_{\bb{r}_-} &= \lambda_\bb{r}\varphi_{\bb{r}_-} - M_\bb{r}\varphi_{\bb{p}}\varphi_0^*.
\end{align}
\end{subequations}
The eigenvalues are quickly found: $\lambda = (\gamma_\bb{r}  -\ji \omega_\bb{r})$ and
\begin{align}
\lambda_\pm = \frac{1}{2}\Big[\lambda_\bb{p} +\lambda_\bb{r} \pm \Big((\lambda_\bb{p} - \lambda_\bb{r})^2 - 8  |\varphi_0|^2 M_\bb{p}M_\bb{r}\Big)^{1/2}\Big].
\end{align}
The destabilizing root is the positive branch which has the real component
\begin{align}\label{posbranch}
\rr{Re}(\lambda_+ ) = \frac{1}{2}\Big[\gamma_+ +\sqrt{\frac{1}{2}} \left(\Omega + \sqrt{\Omega^2+\Theta^2}\right)^{1/2}\Big],
\end{align}
where
\begin{align}
   \Omega&\doteq \gamma_-^2 - \omega_-^2 - 8  |\varphi_0|^2(M_\bb{p}^{\rr{Re}}M_\bb{r}^\rr{Re}- M_\bb{p}^{\rr{Im}}M_\bb{r}^\rr{Im}),
\\ \Theta &\doteq 2\omega_- \gamma_- + 8|\varphi_0|^2 (M_\bb{p}^\rr{Re}M_\bb{r}^\rr{Im}+M_\bb{p}^\rr{Im}M_\bb{r}^\rr{Re}).
\end{align}
Immediately, certain terms appear with physical importance. For instance, $\omega_-$ represents the modulational  part of the dispersion relation that is due to dispersive effects, while $\gamma_+$ is the coupling of the linear growth rates by the nonlinear interaction.
Equation~\eq{posbranch} contains a wealth of information that must be carefully parsed. 

%\subsubsection{Maximally-Coupled Modes}

When the outer discriminant of (\ref{posbranch}) vanishes, the real part of both the unstable and stable eigenvalues becomes $\rr{Re}(\lambda_+) = \gamma_+/2$. This corresponds to a special situation where both the drift-wave and sideband modes become maximally coupled, and occurs when $\Theta = 0$ and $\Omega \le 0$. Here, both modes work together in tandem to create a coupled mode with an effective damping rate that is the average of the individual growth rates. This coupled mode will be referred to as the Maximally-Coupled Mode (MCM). In principle, MCMs can exist for arbitrary drift-wave/sideband pairings where the primary drift-wave mode contains a non-zero radial component $p_x$, though in this section I only consider the most unstable mode with $p_x = 0$. 

The upper bound of the Dimits shift for the 4MT can then be quickly found as the solution to the equation $\gamma_+ = 0$, or
\begin{equation}\label{DSana}
\beta^* = \frac{(D_\bb{p} + D_\bb{r})}{p_y}\frac{[(1+p_y^2)^2 + \THd_\bb{p}^2] [(1+q_x^2 + p_y^2)^2 + \THd_\bb{r}^2]}{\THd_\bb{r}[(1+p_y^2)^2 + \THd_\bb{p}^2] + \THd_\bb{p}[(1+q_x^2 + p_y^2)^2   + \THd_\bb{r}^2]  }.
\end{equation}
This predicts the end of the Dimits shift for the system considered in Sec.~\ref{dns} to be $\beta^* = 5.2$, in excellent agreement with the simulation result in figure~\ref{fluxes}.

%\subsubsection{The most stable zonal amplitude}
 
 In order to develop MCMs, the condition $\Omega \le 0$ must be satisfied. In order for zonal flows to be stabilizing, the condition
\begin{equation}
M_\bb{p}^{\rr{Re}}M_\bb{r}^\rr{Re}- M_\bb{p}^{\rr{Im}}M_\bb{r}^\rr{Im} > 0
\end{equation}
must be met.
It is important to realize that $M_\bb{k} \sim q_x p_y$ times a factor of order unity. This term then clearly represents a Kelvin-Helmholtz-type destabilization  of the zonal mode, and can become dominant for sufficiently large zonal amplitudes and negative values of the mode-coupling coefficient. However, this term only needs to be  comparable to $\gamma_-^2 - \omega_-^2$ in order to spoil the coupling between drift-wave modes. This happens for a sufficiently large value of $q_x$ (denoted by $q_x^*$) and signifies the smallest scale where the last MCM is formed.

 The other condition that must be satisfied in order for MCMs to develop is $\Theta =0$.
There then exists a zonal amplitude that is most stable, given by
\begin{equation}
|\varphi_0|^2 = -\frac{\omega_- \gamma_-}{4 (M_\bb{p}^\rr{Re}M_\bb{r}^\rr{Im}+M_\bb{p}^\rr{Im}M_\bb{r}^\rr{Re})}.
\end{equation}
The stability of such a mode has been verified numerically. 
A few special cases are noted.
First, if $M_\bb{p}^\rr{Im} = M_\bb{r}^\rr{Im}= 0$ and $\omega_- \ne 0$ (as would be the case if linear instability was introduced by hand to the (m)HME without modifying the Poisson equation) then maximal coupling can only be achieved with $\varphi_0 \rightarrow \infty$, though typically the frequency mismatch $\omega_-$ is quite small.
Second, if  $M_\bb{p}^\rr{Im} =M_\bb{r}^\rr{Im} = \omega_- = 0$, then any amount of zonal amplitude is either stabilizing or destabilizing, depending on the sign of $M_\bb{p}^{\rr{Re}}M_\bb{r}^\rr{Re}- M_\bb{p}^{\rr{Im}}M_\bb{r}^\rr{Im}$. Once the discriminant becomes purely imaginary, additional zonal amplitude ceases to make any further difference in terms of stability. This happens when $\Omega = 0$.
Finally, in the (m)HME limit, $\Theta = 0$ and $\Omega = - \omega_-^2 - 8 |\varphi_0 |^2 M_\bb{p}^{\rr{Re}}M_\bb{r}^\rr{Re}$.  Clearly, the effect of frequency mismatch between modes is a stabilizing influence, and is related to the requirement of resonant triads in the theory of wave turbulence where the size of the nonlinearity is asymptotically small compared to the linear terms.

In the introduction, four ingredients in a successful model of the Dimits shift were highlighted, one of which was the necessity for a nonlinear boost in the interaction between zonal flows and drift waves. This boost is provided by the $\h{\alpha}$ operator, which modifies the ion density in both the partial derivative and the Poisson bracket of the HME. While the former effect has been appreciated in the literature~\citep{diamond2005}, the latter  has largely gone unnoticed, and in some cases is entirely neglected~(e.g., \citealt{Holland,Makwana1}). 
To appreciate the importance of the $\bb{E}\btimes\bb{B}$ nonlinearity, consider the mode-coupling coefficients of the (m)HME where $M_\bb{p}^{\rr{Im}}M_\bb{r}^\rr{Im} = 0$ (equation \eq{mcc} with $\THd_\bb{k} =0$) with a background drift-wave at wavenumber $p_y$. In order for energy to flow from an unstable drift-wave mode to a stable one through a zonal flow, the zonal flow itself must be stable to the Kelvin-Helmholtz instability. Otherwise, energy is fed back from the zonal flow to the unstable drift-wave mode. For a zonal flow to be stabilizing, the condition $M_\bb{p}^{\rr{Re}}M_\bb{r}^\rr{Re} >  0$ must be met, leading to destabilizing zonal flows at scales given by
\begin{equation}
q_x > p_y
\end{equation}
 for the HME ($\h{\alpha}_\bb{q}  =1 $). This is equivalent to the stability threshold given by~\citet{diamond2005}, where the hydrodynamic limit was considered.
However, the $\h{\alpha}$ operator in the mHME (viz. \eq{mcc} with $\h{\alpha}_\bb{q}  = 0 $)  leads to zonal flows that are stable against the Kelvin-Helmholtz instability down to the scale set by
\begin{equation}
q_x^{2} = 1 + p_y^{2}.
\end{equation}
Typically, $p_y \ll 1$ in toroidal plasmas, leading to $q_x$ and $p_y$ that are of disparate scales.  This is a rather surprising result, as stable zonal flows at scales on the order $\sim\rho_\rr{s}$ can be an efficient catalyst of \emph{nonlocal} energy transfer, which is in stark contrast to the scale-by-scale transfer in a Kolmogorov-type cascade, whose transfer rate is set by the local eddy turnover time. Additionally, the magnitude of $ M_\bb{p}^{\rr{Re}}M_\bb{r}^\rr{Re}$ in the mHME is increased by a factor of $p_y^{-2}$ for $p_y \ll 1$ compared to that of the HME, resulting in a much stronger stabilizing effect. 
\footnote{This, however, comes at the price of an increased \emph{destabilizing} effect by the Kelvin-Helmholtz instability for scales with wavenumber $q_x^2 \gg 1 + p_y^2$. Luckily, the amplitudes of the zonal flows at these scales tend not to be large.}

\subsection{Heuristic calculation of the Dimits shift}\label{DScalc}

I now generalize the above results for the 4MT in order to heuristically calculate the Dimits shift for the fully nonlinear system. One plausible way to do so is to assume that the zonal interaction couples the most unstable drift-wave mode to \emph{every} sideband mode within the interaction range given in the previous section. Then the threshold for stability is not given by $\gamma_+ = 0$, but rather $(2 \jpi / L_x)\sum_{\rr{MCM}} \gamma_\bb{k} \approx \int_0^{q_x^*} \gamma_\bb{k}(q',p_y^*)  \dd q'= 0$, where the sum is over all modes with $k_y = p_y^*$ in the range $|k_x| \le q_x^*$. Here, I have used the idea of phase space contraction that was given in section~\ref{4mtsec}, but now only consider modes that are well-coupled by zonal shearing to the primary (most unstable) drift-wave mode.
Passing into the continuum here is equivalent to taking the large box limit ($L_x \rightarrow \infty$).

The above calculation differs fundamentally from the linear stability analyses of both~\citet{rogers} and~\citet{numata}. While the latter set of calculations focus on the stability of simplified background zonal flow profiles (resulting in localized instabilities), here the details of the background profile are unimportant; rather, the background profile is assumed to be whatever it needs to be in order to maximally couple the primary drift-wave mode to stable ones, and that it is such that Kelvin-Helmholtz modes are not excited. The criteria for stability then is not determined by the flow profile, but rather by the zonal flow's ability to transfer enough energy to quench the primary instability. Thus, the stability of only a small subset of modes needs to be considered, those that are maximally coupled to the primary drift wave.

The physical picture is as follows: as zonal flows are generated from the secondary instability, drift-waves become sheared, resulting in a direct transfer of energy  to smaller scales. As energy flow is principally horizontal in $k$~space (i.e. energy is not transferred between bands with different $k_y$), the relevant value of $p_y$ that determines stability is that given by the most unstable drift-wave mode, denoted by $p_y^*$. This transfer can be done efficiently to a cluster of modes down to the scale given by $q_x^*$, where the effect of zonal flows ceases to be stabilizing. Thus the critical gradient that signifies the end of Dimits shift will be roughly given when this cluster of modes goes unstable (becomes volume expanding, see~\citealt{TerryHorton2}).

Quantitatively, this situation can be described as a system of four equations,
\begin{subequations}\label{thething}
\begin{align}
 \left.\frac{\partial \gamma_{\bb{p}}}{\partial p_y}\right|_{p_y = p_y^*} &= 0,
 \\ \int_0^{q_x^*} \gamma(q', p_y^*) \dd q' &= 0,
\\ \Theta &= 0,
\\ \Omega &= 0,
\end{align}
\end{subequations}
and four unknowns $q_x^*$, $p_y^*$,  $|\varphi_0|^2$, and $\beta^*$.
The first equation relates  $p_y^*$ of the most unstable drift-wave mode to the instability parameter $\beta^*$. 
The second equation determines the critical gradient where the cluster of coupled modes becomes unstable (volume expanding), and relates $q_x^*$,  $p_y^*$, and $\beta^*$.  
Finally, the last two equations leverage the calculations of the 4MT to approximately determine the range at which MCMs can be formed, and relates $|\varphi_0|^2$, $q_x^*$,  $p_y^*$, and $\beta^*$. As a somewhat rougher  approximation, one can alternatively use  $q_x^{*2} = 1 + p_y^{*2}$ in place of (\ref{thething}c--d), eliminating the need of $|\varphi_0|^2$. Finally, for systems where the growth rates are rather difficult to come by, one could also use the approximate condition $\gamma(0,p_y^*)+ \gamma(q_x^*,p_y^*)= 0$, where the coupling between only the most unstable mode and the sideband at scale $k_x = q_x^*$ is considered. While this forgoes the phase-space argument of~\citet{TerryHorton2} (see section~\ref{4mtsec}), it does provide a simple and reasonable estimate of the size of the shift.

Numerical solution of (\ref{thething}a--d) with $\THd_\bb{k} = 1.5 k_y$ and $D = 1 - 0.01 \nabla_\perp^2$
yields $\beta^* \approx 5.79$ and results in a Dimits shift size of $\Delta \beta \approx 1.55$.
This is within a percent of the value obtained by the direct numerical simulation in Sec.~\ref{dns} ($\beta^* \gtrsim 5.75$ and $\Delta \beta \approx 1.5$). Using the same methodology, the calculated shift for the system parameters given in the caption of figure~\ref{time} is roughly $\beta^* \approx 6.77$. While the Dimits shift for this system is infinite, this agrees well with the end of the first regime at $\beta \approx 7$.
These are fairly good estimates, considering they are the result of a straightforward linear calculation with only a few nonlinear considerations. Notice that this analysis is insensitive to the details of the underlying model; it applies equally well to \emph{any} Hasegawa-Mima-like equation, and should be generalizable to more complicated ones, provided they retain the $\bb{E}\btimes\bb{B}$ nonlinearity. Additionally, it also works with any generic damping operator. 
I stress that this is only an estimate; it is neither an upper bound, as additional avenues of efficient energy transfer may exist, nor is it a lower bound, as the existence of a stable state does not guarantee it is physically realizable.

\begin{figure}
\centering
\subfloat[$D_\bb{k} =  1 + 0.01 k_\perp^2$]{
\includegraphics[scale=0.80]{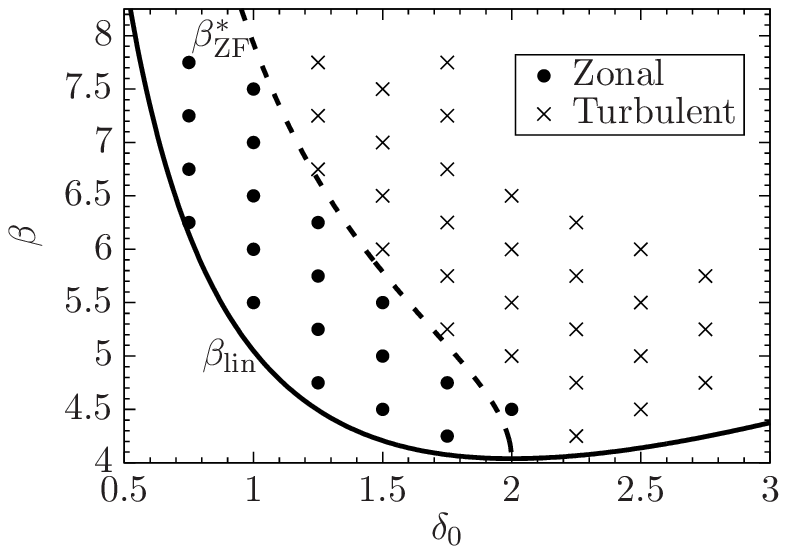}
}
\subfloat[$D_\bb{k} = 0.3 |k_y|$]{
\includegraphics[scale=0.80]{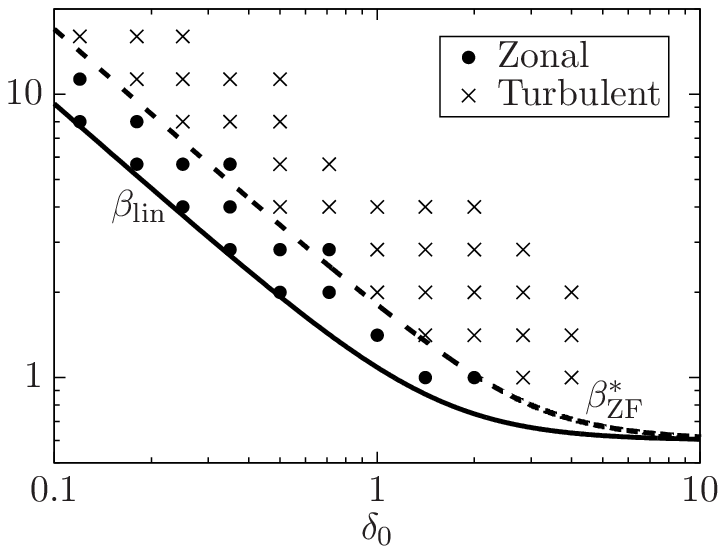}
}
\caption{\label{scan} Parameter scan of the nonlinear system with  $\THd_\bb{k}$ given by \eq{deltak2}. The left panel (a) uses $D_\bb{k} = 1 + 0.01 k_\perp^2$, while the right panel (b) uses $D_\bb{k} = 0.3 |k_y|$.
Bold dots denote system that end in steady zonal states while crosses denote systems that end in turbulent states.  The solid line marks the linear stability threshold, while the dashed line denotes the predicted end of the Dimits shift $\beta^*$ given from the solution of \eq{thething}.}
\end{figure}

Figure~\ref{scan} shows a parameter scan of the system with $\THd_\bb{k} = \ji \delta_0 k_y$ over  both $\delta_0$ and $\beta$. The left panel (a) uses frictional and viscous damping with $D_\bb{k} = 1 + 0.01 k_\perp^2$, while the right panel (b) uses a damping operator appropriate for a Landau-fluid closure with $D_\bb{k} = 0.3 |k_y|$. These simulations are run for a maximum time of $t = 10\,000$.
Bold dots denote simulations that end in steady zonal states while crosses denote systems that end with turbulence. The solid line shows the linear stability threshold while the dashed line shows the threshold calculated from (\ref{thething}a--d). This plot shows excellent agreement between the estimation and the observed numerical boundary of the Dimits shift. Surprisingly, in panel (a) the estimate also reasonably predicts the boundary at $\delta_0 = 2$. 

One reason why this heuristic calculation works well for this system is that there are only two competing mechanisms for energy transfer: the direct cascade (which is local), and the zonal shearing (which is nonlocal). While the nonlocal mechanism can be taken into account in more complete systems,  added complexity comes with additional competing forms of energy transfer. One such mechanism is the dual cascade, which has the ability to transfer both energy to large scales (where large-scale sinks may reside) and enstrophy \emph{nonlocally} to small scales. This process is dominated by the largest eddies of the system, and underlies the typical picture of shearing by large-scale zonal flows~\citep{diamond2005}.  More complex models also have an increased number of fields, leading to multiple branches (see the recent work of \citealt{Makwana1,Makwana2}). Typically, one branch is unstable in a region of $k$-space while the others are purely stable for all $k$. Coupling energy between branches, then, can lead to an efficient source of energy dissipation.
This is exacerbated when moving to a kinetic system, where there could be an arbitrary number of stable branches.  Finally, additional nonlinearities can also arise, such as finite Larmor-radius terms. Thus, one has to take into account all these effects as they arise, and determine their relative impact on stability accordingly.

\section{Discussion and Conclusion} \label{discussion}

I have shown through direct numerical simulation that the Terry-Horton equation can be made to exhibit the Dimits shift with two suitable modifications. First, an appropriate adiabatic electron response is added to ensure that electrons do not respond to a potential that is constant along a flux surface. Secondly, zonal modes are made to be explicitly undamped, thus capturing the residual Rosenbluth-Hinton states seen in gyrokinetic simulations. The Dimits shift is shown to persist even after various simplifications are made. Analytical progress was made on a four-mode truncation of the system, focusing on the behaviour of the zonal mode growth during the secondary instability, and calculating an upper bound on the end of the Dimits shift.

Using this model, important insight was gained on the underlying mechanism by which the Dimits shift operates. Specifically, it was shown that the inclusion of the appropriate adiabatic electron response results in a $\bb{E}\btimes\bb{B}$ nonlinearity that can transfer energy \emph{nonlocally} from large scales to scales of order $\sim \rho_\rr{s}$. The nonlocal transfer as a result of nonadiabatic electron response in the $\bb{E}\btimes\bb{B}$ nonlinearity has been known for some time~\citep{Liang1993}. However, while the nonlocal contribution from \emph{adiabatic} electrons in the $\bb{E}\btimes\bb{B}$ nonlinearity has been appreciated in some areas of the literature (see, for example,~\citealt{rogers,guzdar}), it has gone unnoticed in other areas. As an example, two studies \citep{Holland,Makwana1} that used a two-field gyrofluid model of the ITG instability neglected the $\bb{E}\btimes\bb{B}$ nonlinearity altogether, and thus this effect was entirely absent. Similarly, the nonlinear contribution from proper adiabatic electron response is not mentioned in the work of~\citet{diamond2005, itoh_2006}, while the linear contribution under the partial time derivative is considered.
 An important implication of this correction is that it results in a broader range of scales where zonal flows are stable to the Kelvin-Helmholtz instability, thus leading to a more favourable stability criterion than that given in the review article by~\citet{diamond2005}, where the hydrodynamic limit was considered.  It is important to realize that, as modes arising from the primary and secondary instabilities are usually at scales larger than the sound radius, the adiabatic $\bb{E}\btimes\bb{B}$ nonlinearity will at least be as large as, if not larger than, the ion-polarization nonlinearity, and so it is never appropriate to neglect this term in the study of ITG modes.

It is worthwhile to clarify a result given in the work of~\citet{rogers}. There, it was stated the the Kelvin-Helmholtz instability is absent in the tertiary instability, and a finite component of $T_\perp$ was needed for instability.
 As an example, they considered the stability of a zonal flow with $k_x \rho_\rr{s} = 0.25$, which was shown to be unconditionally stable to the Kelvin-Helmholtz instability (consistent with the analysis of section~\ref{tertiary}.) However, it must be understood that modes with $k_x \rho_\rr{s} > 1$ \emph{can} go unstable to the Kelvin-Helmholtz instability, even without a component of $T_\perp$. While these small scale modes are not relevant when discussing the usual ion-polarization-driven forward cascade which is dominated by zonal flows at the largest scales, it is rather crucial for the behaviour of the adiabatic $\bb{E}\btimes\bb{B}$ nonlinearity, which depends on zonal flows at a broad range of scales. At which scales these flows ultimately go unstable, then, determines the effective interaction range in $k$-space of this nonlinearity. 

It is also stated in the work of~\citet{rogers} that the addition of background gradients generally overpowers the instability given by $T_\perp$. This is interesting in the context of equation~\ref{posbranch}, which states that (at least in the four-mode truncation of the mTHE) the stability criterion regarding the end of the Dimits shift is ultimately dictated by the linear terms, regardless of the size of the zonal flow. Thus, neglecting the linear terms in a stability analysis of the Dimits shift regime~\emph{may never be justified}, even when other avenues of destabilization are present.

Future work should focus on extending this analysis to more physically complete (e.g., gyrokinetic) systems to see if one can derive the size of shift originally seen for the Cyclone Base Case in~\cite{Dimits}. Other quantities of interest are the saturation levels and spectra of the zonal flows resulting from the secondary instability, and the saturation levels of turbulent transport beyond the shift. While this work was built upon a very idealized model, the $\bb{E}\btimes\bb{B}$ nonlinearity that arises from the non-adiabatic electron response to zonal modes is present in most relevant models of tokamak plasmas, including gyrokinetics and higher order gyrofluid closures. The nonlocal transfer of energy catalyzed by zonal flows, then, is an effect that should remain important in more complete systems. Future research should aim to confirm this.

Finally, a better estimate of the size of the Dimits shift for the mTHE could be derived using more rigorous methods (at the level of~\citealt{krommes_kim}).  While it is known that  the energy transfer at large scales is dominantly to zonal models, if this can be shown to be the case down to scales with wavenumber $q_x$, defined using $ q_x^2 \rho_\rr{s}^2\sim 1 +  p_y^{2} \rho_\rr{s}^2$  where $p_y$ is the wavenumber of the most unstable mode, then the heuristic calculations presented here could be better justified.
This can be done, in principle, by using
statistical closures to study the nonlinear mechanisms.  However, as~\citet{Parker2} have pointed out,
it is necessary to begin with an inhomogeneous closure in order that one
can consider inhomogeneous symmetry-breaking perturbations (zonal flows) to
a state of homogeneous turbulence.  (As was discussed by \citealt{StOnge1}, here `turbulence' below the point of zonostrophic
instability refers to homogeneous noise due to discrete particles.)  Not
only does an inhomogeneous closure allow for symmetry breaking, it contains
all of the physical effects involved in destabilizing those flows and
allowing for a transition from the Dimits-shift regime to states of fully
developed turbulence.  Because the general structure of an inhomogeneous
closure is necessarily complicated, carrying out such a program to
completion represents a significant challenge for the future.

\acknowledgments
I would like to thank J. A. Krommes, G. W. Hammett, and T. Stoltzfus-Dueck  for many instructive discussions. I would also like to thank J. B. Parker and J. Squire for help with the numerics, and M. W. Kunz for his steadfast and unwavering support of this research.  I would like to thank the anonymous referees for constructive comments that led to a much improved presentation. Finally, I would like to thank the editor B. Dorland for offering additional background and historical context on the subject of this paper.
This work was supported by U.S. DoE contract DE-AC02-09CH11466.

\bibliography{refs}

\end{document}